\documentclass{emulateapj}
\usepackage{graphicx,amssymb}
\usepackage[usenames,dvips]{color}

\newcommand{\kms}{\ifmmode {\rm km\,s}^{-1} \else km\,s$^{-1}$\fi}

\shorttitle{Monitoring of 1720 MHz OH masers}
\shortauthors{Pihlstr\"om et al.}

\begin{document}

\title{Very Large Array monitoring of 1720 MH\lowercase{z} OH masers toward the
  Galactic center}

\author{Y.~M.~Pihlstr\"om\altaffilmark{1}}\affil{Department of Physics
  and Astronomy, University of New Mexico, MSC07 4220, Albuquerque, NM
  87131}

\email{ylva@unm.edu}

\altaffiltext{1}{Y.~M.~Pihlstr\"om is also an Adjunct Astronomer at the
  National Radio Astronomy Observatory}

\author{L.~O.~Sjouwerman}
\affil{National Radio Astronomy Observatory, P.O. Box 0, Lopezville
  Rd.\,1001, Socorro, NM 87801}

\and

\author{R.~A.~Mesler} \affil{Department of Physics and Astronomy,
  University of New Mexico, MSC07 4220, Albuquerque, NM 87131}

\begin{abstract}
  We present the first variability study of the 1720 MHz OH masers
  located in the Galactic Center. Most of these masers are associated
  with the interaction between the supernova remnant Sgr\,A\,East and
  the interstellar medium, but a few masers are associated with the
  Circumnuclear Disk (CND). The monitoring program covered five epochs
  and a timescale of 20-195 days, during which no masers disappeared
  and no new masers appeared. All masers have previously been detected
  in a single epoch observation about one year prior to the start of
  the monitoring experiment, implying relatively stable conditions for
  the 1720~MHz OH masers. No extreme variability was detected.  The
  masers associated with the northeastern interaction region between
  the supernova remnant and the $+50$~\kms\ molecular cloud show the
  highest level of variability. This can be explained with the
  $+50$~\kms\ molecular cloud being located behind the supernova
  remnant and with a region of high OH absorbing column
  density along the line of sight. Possibly the supernova remnant
  provides additional turbulence to the gas in this region, through
  which the maser emission must travel.  The masers in the southern
  interaction region are located on the outermost edge of Sgr\,A\,East
  which line of sight is not covered by either absorbing OH gas or a
  supernova remnant, in agreement with the much lower variability
  level observed. Similarly, the masers associated with the CND show
  little variability, consistent with them arising through collisions
  between relatively large clumps of gas in the CND and no significant
  amount of turbulent gas along the line of sight.
\end{abstract}


\section{Introduction}
Masers observed in the 1720 MHz satellite line of OH are considered
signposts of regions of shocked gas, since they often are associated
with supernova remnants (SNRs) \citep[e.g.,][]{frail94,green97}. The
1720\,MHz masers are not exclusively found in SNRs but are also
detected in star forming regions
\citep[e.g.,][]{macleod97,szymczak04,niezurawska04}, demonstrating
that 1720\,MHz masers may occur under different pumping conditions and
OH column densities than those present in SNRs
\citep{gray91,gray92}. Concentrating on the SNRs, the masers originate
in the shocked region where the expanding SNR collides with the
surrounding interstellar medium (ISM). Consistent with this model, a
large number of 1720\,MHz masers are found in the Galactic center (GC)
where the Sgr\,A\,East SNR plows into the ISM, notably into the
$+50$~\kms\ molecular cloud \citep[e.g.,][hereafter
Paper\,I]{sjouwerman08}. The majority of these masers are observed
along a circular pattern outlined by the expansion of the SNR,
displaying a relatively small range of line of sight velocities
between $+34\leq V_{\rm LSR}\leq 66$\,km\,s$^{-1}$
\citep[e.g.,][]{yusefzadeh96,karlsson03,sjouwerman08}. In addition,
two separate groups of masers are located near the Circumnuclear Disk
(CND) that are not directly explained by the SNR/ISM interaction
model. These two groups of masers have velocities that are offset from
the Sgr\,A\,East masers; $V_{\rm LSR}\simeq+130$\,km\,s$^{-1}$ and
$V_{\rm LSR}\simeq-130$\,km\,s$^{-1}$ respectively (Paper\,I). In
Paper\,I we argue that these masers are unlikely to be pumped by a
shock produced by Sgr\,A\,East. Other plausible pumping scenarios
include local shocks produced by random motions of clumps or
turbulence, supported by the presence of strong H$_2$ (1-0) $S(1)$
emission \citep{yusefzadeh01}, or by infrared (IR) pumping similar to
conditions observed in star forming regions \citep{gray92}.

The environment in the CND is likely to be different from that in a
SNR/ISM post-shock region. Variability studies of masers can be used
to probe the environment and could further shed light on the
differences in excitation mechanisms and conditions for CND versus
SNR/ISM masers. In the CND, the pumping may include IR pumping routes,
since local IR peaks are observed within the CND
\citep{latvakoski99}. This would be in contrast to the 'standard' SNR
masers, which are pumped by collisions only. Also, differences in
collision rate along the line of sight, due to clumpiness of the
medium, could result in a different maser flux variability of the CND
masers as compared to the SNR/ISM masers
\citep{pihlstrom08,pihlstrom01}.

\begin{deluxetable}{rlccc}[th]
\tabletypesize{\scriptsize}
\tablecaption{Observational details of the VLA monitoring program performed 
in 2006}
\tablewidth{0pt}

\tablehead{ \colhead{$\phantom{0000}$Day of} & \colhead{Config.} & \colhead{Beam size} & \colhead{Channel rms} & \colhead{Obs.\ time}\\
  \colhead{$\phantom{0000}$Year} & & \colhead{\arcsec$\times$\arcsec} & \colhead{mJy\,beam$^{-1}$} & \colhead{hours}
}
\startdata                                                                      
1$\phantom{0000}$47 & A   & 2.7$\times$1.2 & 5.2 & 5 \\
2$\phantom{000}$144 & BnA & 3.9$\times$2.8 & 6.3 & 4 \\  
3$\phantom{000}$164 & BnA & 5.7$\times$3.3 & 7.2 & 4 \\
4$\phantom{000}$196 & B   & 7.7$\times$3.8 & 8.0 & 4 \\
5$\phantom{000}$242 & B   & 7.6$\times$3.4 & 6.6 & 4 \\
\enddata                                                                        
\label{observations}                                                            
\end{deluxetable}

Variability of other OH maser transitions has been observed. For
example, the 1612, 1665 and 1667 MHz circumstellar OH masers in
variable stars frequently exhibit variability that is correlated with
the stellar cycle \citep[e.g.,][]{jewell79, etoka00}. Thus, this
variability can be directly coupled to the infrared pumping mechanism
of 1612, 1665 and 1667 MHz masers. For 1720 MHz masers in star forming
regions, variability over two epochs have been reported by
\citet{caswell04}, and variability on 15-20 minute timescale is seen
for the masers in the star forming region W3(OH)
\citep{ramachandran06}. However, the pumping of 1720 MHz masers in
star forming regions include radiative pumping routes via IR emission
from the central star \citep{gray91, gray92}. Such radiative pumping
scheme is not available in the SNRs, which are completely
collisionally pumped \citep{elitzur76, lockett99, wardle99,
  pihlstrom08}. Since these masers are excited by the SNR shock, it is
less likely that any observed variability would be induced by
pumping variations. Instead we expect that variations in the
amplification path length would be the most direct method to produce
amplitude variations in SNR/ISM 1720 MHz masers.

As little is known about the 1720 MHz maser variability in SNRs, we
here present a five-epoch variability study of the 1720 MHz masers in
the GC with the Very Large Array (VLA).

\section{Observations}\label{obs}

During 2006, five observations of the 1720\,MHz OH masers in the GC
were performed using the VLA under the project code AP500. Each
observation had an observing time of typically four hours and used a
single pointing centered on Sgr\,A* (J2000 coordinates RA 17 45
40.038, Dec $-$29 00 28.07). Observations were made over a total
period of 195 days, with successive irregular intervals ranging from
20 to 97 days (Table \ref{observations}). The observations were done
with dual circular polarization using two IF pairs of 1.562\,MHz
bandwidth each. The two IFs were centered at offset velocities in
order to cover a large, total velocity range
($-$243\,km\,s$^{-1}<V_{\rm LSR}<245$\,km\,s$^{-1}$). With 128
channels and no on-line smoothing applied, the resulting velocity
resolution was 2.6\,km\,s$^{-1}$. The data were calibrated using
VLARUN, a pipeline VLA data reduction procedure available in
AIPS. After continuum subtraction in the {\it u,v}--plane the data was
imaged with natural weighting using standard AIPS
procedures. Phase-only self-calibration was performed against the
brightest maser feature with a peak flux density of 5.2\,Jy at 66
km\,s$^{-1}$. Full spectral line cubes were made using the AIPS task
IMAGR. Depending on the VLA configuration, the resulting data cubes
had slightly different synthesized beam sizes, listed in Table
\ref{observations} along with the typical rms channel noise in each
cube. All data was averaged over the length of the observing
run. Shorter averaging times are not considered in this paper since
the increased signal to noise ratio would prevent many masers from
being reliably detected.

The amplitude scale was calibrated using 3C286 as the absolute flux
calibrator, and then bootstrapped to the phase calibrator
1751$-$253. The mean flux density of 1751$-$253 over Epoch 1-5 was 998
mJy and varied less than 3\%. The flux density bootstrapping procedure
resulted in average amplitude corrections of 1.03$\pm$0.09,
corresponding to an amplitude uncertainty of 9\%, thus dominating the
flux errors.

To identify masers, the AIPS task SQASH was used to produce maps of
the maximum intensity over the velocity axis, for each sky pixel. In
each of those resulting 'maxmaps' \citep{sjouwerman98}, a feature was
considered a detection if the flux density exceeded 8$\sigma$. A
weaker feature can be considered a detection, if it occurred at a
position already known to harbor a maser (e.g., in Paper\,I), or if it
would occur in at least three of the five epochs. We acknowledge that
this potentially would bias toward non-(diss)appearing sources, but we
have not found this to be a problem in this data. To calculate the
integrated flux given in Table \ref{fluxes} channels with emission
down to three times the rms noise were included.

\section{Results}
\label{results}

We use same notation of masers previously detected as in Paper\,I, in
order of increasing RA (see Table \ref{fluxes}). In total 27 masers
were detected, out of which one was new\footnote{This maser is among a
  few potential detections that did not meet the strict 10$\sigma$
  detection criteria in Paper\,I (it was a 7.8$\sigma$ peak detection
  there, consistent with $VI>1$).} (maser ``Y'').  Maser $7+9$ are
listed as a single maser, as it was unresolved and seen as a single
maser in the last epoch. Table \ref{fluxes} presents the properties of
each maser feature in each epoch, including velocity integrated flux,
the number of channels with flux density exceeding 3 times the rms
noise, the variability index $VI$ and the maser region. We define $VI$ as:

\begin{equation}
  VI=\frac{1}{n}\sum{\left|\frac{S_i-\bar{S}}{\bar{\sigma}}\right|}
\end{equation}

where $n$ is the number of epochs, $S_i$ the integrated flux at epoch
$i$, $\bar{S}$ the average integrated flux and where $\bar{\sigma}$ is
the average flux error over all epochs. This variability index is a
measure of average variability about the mean, in units of the rms
flux error. We consider a maser variable if $VI > 1.0$, i.e.\ if the
flux on average varies more than the average error in the flux.

\begin{deluxetable*}{r|rrrrrrrrrr|rc}
  \tabletypesize{\scriptsize} \tablecaption{Maser flux history}
  \tablewidth{0pt}
  \tablehead{\colhead{No} & \multicolumn{10}{c}{Integrated flux\tablenotemark{a} and number of channels\tablenotemark{b}} & \colhead{VI} & \colhead{Region}\\
    &  \multicolumn{10}{c}{ (Jy\,km\,s$^{-1}$)} & }
\startdata
  & \underline{Epoch 1} & \underline{N$_{ch}$} & \underline{Epoch 2} & \underline{N$_{ch}$} & \underline{Epoch 3}&  \underline{N$_{ch}$} & \underline{Epoch 4} & \underline{N$_{ch}$} & \underline{Epoch 5} & \underline{N$_{ch}$} & & \\
  1 &  0.23$\pm$0.03 & 2 & 0.22$\pm$0.03 & 2 & 0.20$\pm$0.03 & 2 & 0.21$\pm$0.03 & 2 & 0.22$\pm$0.03 & 2 & 0.31 & CND\\
  2 &  0.08$\pm$0.02 & 1 & 0.06$\pm$0.02 & 1 & 0.05$\pm$0.02 & 1 & 0.08$\pm$0.03 & 1 & 0.07$\pm$0.02 & 1 & 0.43 & CND \\
  3 &  0.20$\pm$0.03 & 3 & 0.21$\pm$0.03 & 2 & 0.24$\pm$0.04 & 2 & 0.24$\pm$0.04 & 2 & 0.23$\pm$0.04 & 2 & 0.46 & CND \\
  5 &  1.68$\pm$0.21 & 2 & 1.75$\pm$0.22 & 2 & 1.76$\pm$0.22 & 2 & 1.78$\pm$0.22 & 2 & 1.74$\pm$0.22 & 1 & 0.11 & NW  \\
  6\tablenotemark{c} &
       0.22$\pm$0.03 & 3 & 0.23$\pm$0.03 & 2 & 0.27$\pm$0.04 & 3 & 0.20$\pm$0.03 & 2 & 0.18$\pm$0.05 & 1 & 0.75 &  \\
$7+9$\tablenotemark{d}& 
       5.54$\pm$0.35 & 4 & 5.66$\pm$0.28 & 5 & 5.67$\pm$0.29 & 5 & 5.65$\pm$0.29 & 5 & 5.42$\pm$0.34 & 4 & 0.28 & CND \\
  8 &  0.49$\pm$0.05 & 3 & 0.48$\pm$0.04 & 2 & 0.44$\pm$0.06 & 2 & 0.47$\pm$0.05 & 3 & 0.47$\pm$0.06 & 2 & 0.23 & NW \\
  10&  0.93$\pm$0.06 & 4 & 1.10$\pm$0.07 & 4 & 1.04$\pm$0.09 & 3 & 1.17$\pm$0.10 & 4 & 0.96$\pm$0.12 & 2 & 0.87 & S\\
  11&  0.21$\pm$0.03 & 3 & 0.27$\pm$0.04 & 4 & 0.15$\pm$0.03 & 1 & 0.46$\pm$0.04 & 5 & 0.28$\pm$0.03 & 3 & 2.26 & NW \\
  12&  0.24$\pm$0.03 & 2 & 0.25$\pm$0.03 & 2 & 0.23$\pm$0.04 & 2 & 0.24$\pm$0.04 & 2 & 0.27$\pm$0.04 & 2 & 0.33 & S \\
  13&  0.19$\pm$0.03 & 2 & 0.14$\pm$0.03 & 1 & 0.17$\pm$0.05 & 1 & 0.16$\pm$0.04 & 1 & 0.17$\pm$0.05 & 1 & 0.39 & S \\
  14&  0.20$\pm$0.03 & 2 & 0.26$\pm$0.03 & 2 & 0.26$\pm$0.04 & 2 & 0.40$\pm$0.04 & 3 & 0.34$\pm$0.05 & 2 & 1.62 & S \\
  15&  0.78$\pm$0.10 & 2 & 0.83$\pm$0.11 & 2 & 0.79$\pm$0.10 & 2 & 0.83$\pm$0.11 & 2 & 0.85$\pm$0.11 & 2 & 0.25 & S \\
  16&  0.53$\pm$0.07 & 2 & 0.59$\pm$0.08 & 2 & 0.62$\pm$0.08 & 2 & 0.62$\pm$0.08 & 2 & 0.61$\pm$0.08 & 3 & 0.40 & S \\
  17&  1.63$\pm$0.14 & 3 & 1.30$\pm$0.11 & 3 & 1.34$\pm$0.12 & 3 & 1.33$\pm$0.12 & 3 & 1.26$\pm$0.11 & 3 & 0.88 & S \\
  18& 13.82$\pm$0.86 & 4 &14.85$\pm$0.93 & 4 &15.33$\pm$0.96 & 4 & 15.47$\pm$0.97& 4 &14.52$\pm$1.82 & 2 & 0.45 & S \\
  20&  0.20$\pm$0.03 & 2 & 0.37$\pm$0.04 & 3 & 0.36$\pm$0.04 & 3 & 0.27$\pm$0.04 & 2 & 0.31$\pm$0.04 & 2 & 1.38 & NE \\
  21&  0.45$\pm$0.06 & 2 & 0.53$\pm$0.07 & 2 & 0.54$\pm$0.07 & 2 & 0.51$\pm$0.07 & 2 & 0.50$\pm$0.07 & 2 & 0.44 & NE \\
  22&  0.25$\pm$0.03 & 2 & 0.26$\pm$0.03 & 2 & 0.22$\pm$0.06 & 1 & 0.17$\pm$0.05 & 1 & 0.19$\pm$0.05 & 1 & 0.65 & NE \\
  23&  0.36$\pm$0.04 & 3 & 0.39$\pm$0.04 & 2 & 0.36$\pm$0.05 & 2 & 0.40$\pm$0.04 & 3 & 0.52$\pm$0.04 & 4 & 1.09 & NE \\
  24&  0.23$\pm$0.03 & 3 & 0.22$\pm$0.04 & 4 & 0.53$\pm$0.05 & 7 & 0.34$\pm$0.04 & 3 & 0.40$\pm$0.04 & 4 & 2.56 & NE \\
  25&  0.30$\pm$0.03 & 4 & 0.30$\pm$0.04 & 2 & 0.35$\pm$0.04 & 3 & 0.50$\pm$0.05 & 6 & 0.42$\pm$0.04 & 4 & 1.78 & NE \\
  26&  0.14$\pm$0.04 & 1 & 0.33$\pm$0.04 & 3 & 0.35$\pm$0.04 & 4 & 0.38$\pm$0.04 & 3 & 0.34$\pm$0.04 & 3 & 1.79 & NE \\
  27&  0.13$\pm$0.02 & 2 & 0.20$\pm$0.03 & 2 & 0.42$\pm$0.04 & 4 & 0.22$\pm$0.04 & 2 & 0.19$\pm$0.03 & 2 & 2.33 & NE \\
 Y\tablenotemark{e}&
       0.15$\pm$0.02 & 3 & 0.29$\pm$0.04 & 4 & 0.24$\pm$0.03 & 4 & 0.34$\pm$0.04 & 4 & 0.19$\pm$0.03 & 2 & 1.75 & NE\\ 
 28&   0.09$\pm$0.02 & 1 & 0.11$\pm$0.02 & 1 & 0.08$\pm$0.03 & 1 & $<0.02\pm 0.04$&  & 0.18$\pm$0.03 & 2 & 1.96 & NE 
\enddata
\tablenotetext{a}{The error in the integrated flux is
  estimated from the rms noise in the image and the uncertainty in the
  line width which is assumed to be 25$\%$ of the channel width.}
\tablenotetext{b}{The number of channels for which the flux density exceeded 3$\sigma$.}
\tablenotetext{c}{This maser is probably not directly associated with
  the SNR/ISM interaction, see Paper\,I}
\tablenotetext{d}{This maser is actually two, but in Epoch 5 they were
  not resolved and are treated as one feature in this paper}
\tablenotetext{e}{New maser (not listed in Paper\,I) at RA
  = 17 45 50.4, Dec = $-28$ 59 13, $V_{\rm LSR} = 53 $\,km\,s$^{-1}$. }
\label{fluxes}
\end{deluxetable*}

The individual maser positions do not vary significantly between the
epochs (less than the measurement errors; 0.3 arcseconds), and we can
therefore conclude they are the same features being observed in all
epochs. Moreover, the positions and velocities here agree with the
positions and velocities of the masers detected in Paper\,I. In
Paper\,I two masers detected in archival data were included for
completeness, even though they were not detected in the observations:
\#4 and \#19. These two masers were not re-detected in our new
observations and we conclude they are extinct. Overall our results
indicate stable conditions over at least year-long term for the
masers.

For the relatively short timescales and sparse sampling of maser
amplitudes reported on here, it is difficult to find patterns such as
periodicity, a steady flux increase or decrease. Future more densely
sampled surveys of the maser amplitudes over timescales of years will
be needed to address the question of whether variability patterns
exist. In this paper we therefore discuss these masers in terms of
showing variability or not. By a visual inspection of
Fig.\ \ref{plots} and the $VI$ in Table \ref{fluxes}, we find that 10
out of 26 masers show variability with $VI>1.0$. No maser is showing
extreme variability, with the highest observed $VI=2.56$.

\begin{figure*}[th!]
\resizebox{18cm}{!}{\rotatebox{0}{\includegraphics{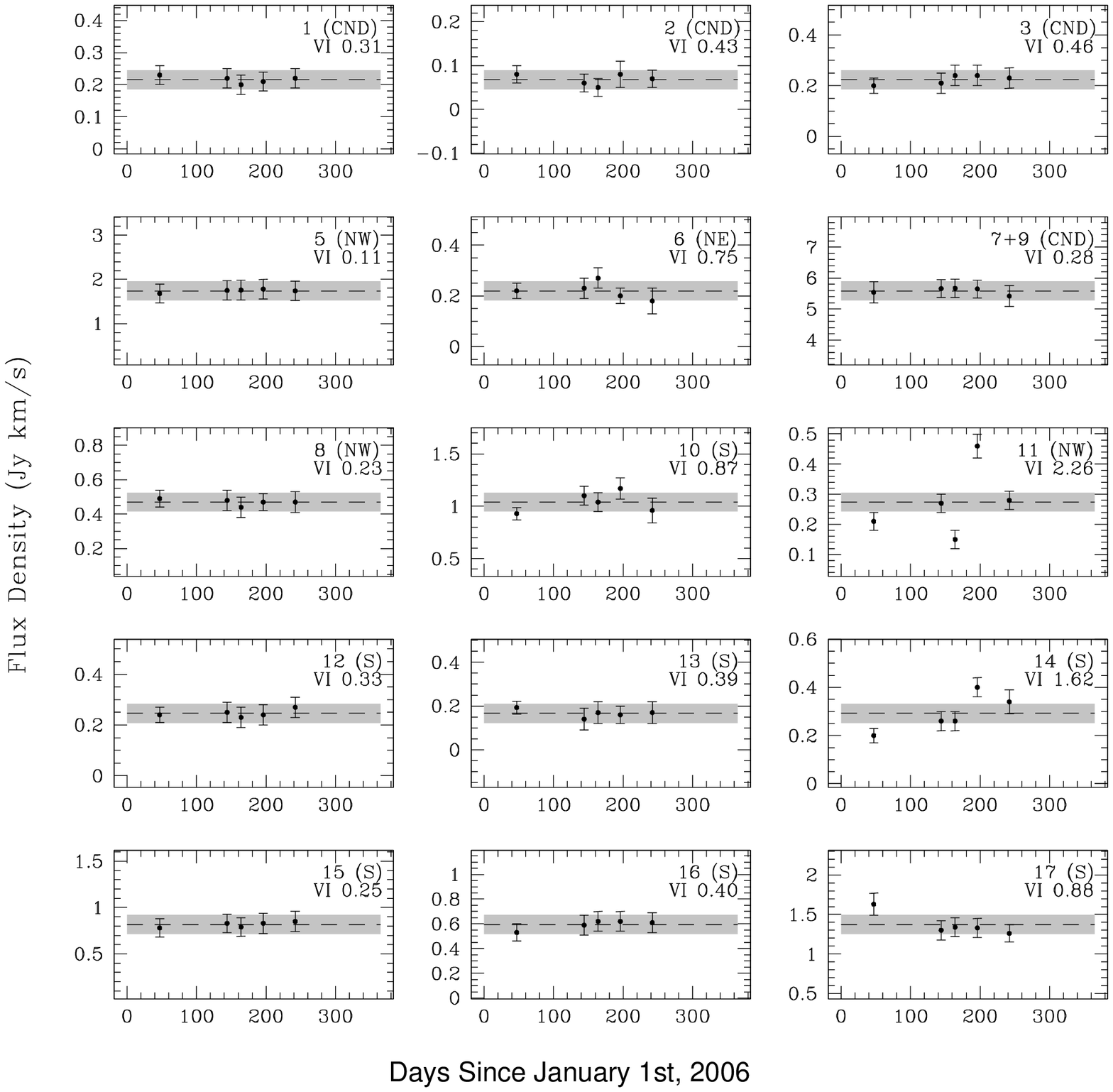}}}
\caption{Velocity integrated flux density versus time, with the x-axis
  showing number of days since January 1st, 2006. The flux density
  measurements are shown with a 1-sigma error bar. The shaded area
  shows the source-average $\pm 1\sigma$ region around the
  source-average flux density value; the flux-range is $\pm
  7\sigma$. Variable masers with $VI>1$ will have one or more
  measurements significantly outside the shaded region. }
\label{plots}
\end{figure*}

\addtocounter{figure}{-1}
\begin{figure*}[th!]
\resizebox{18cm}{!}{\rotatebox{0}{\includegraphics{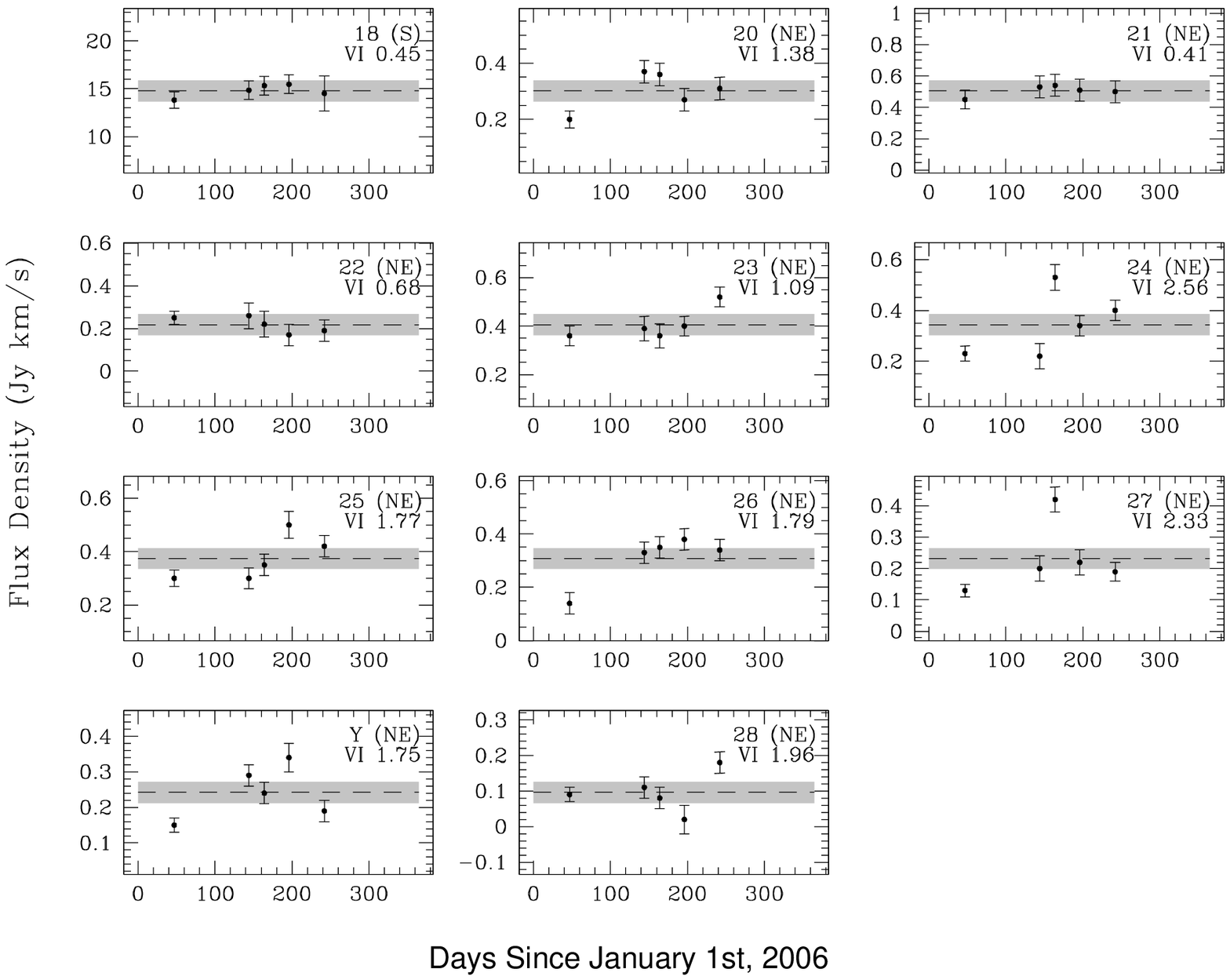}}}
\caption{Continued.}
\end{figure*}

Maybe except for the weakest sources in Paper\,I (\#11, \#20 and \#22)
we do not see a correlation between high $VI$ and weak flux, where
measured flux errors may be a large fraction of the measured flux. We
also do not see a correlation between flux and array configuration,
where larger synthesized beams (at later dates) would be able to pick
up more flux if the masers were angularly extended. There may be a
bias in the flux if the measurements at individual epochs have used
different numbers of channels. The effect is not entirely clear, and
one could argue that varying maser emission, apart from varying in
flux, may also vary in width of the spectral feature.  The results
obtained below seem to indicate that this potential dependence would
only cause a minor adjustment to the $VI$ numbers, but not the trend.

\begin{figure}[th]
\resizebox{8.5cm}{!}{\rotatebox{0}{\includegraphics{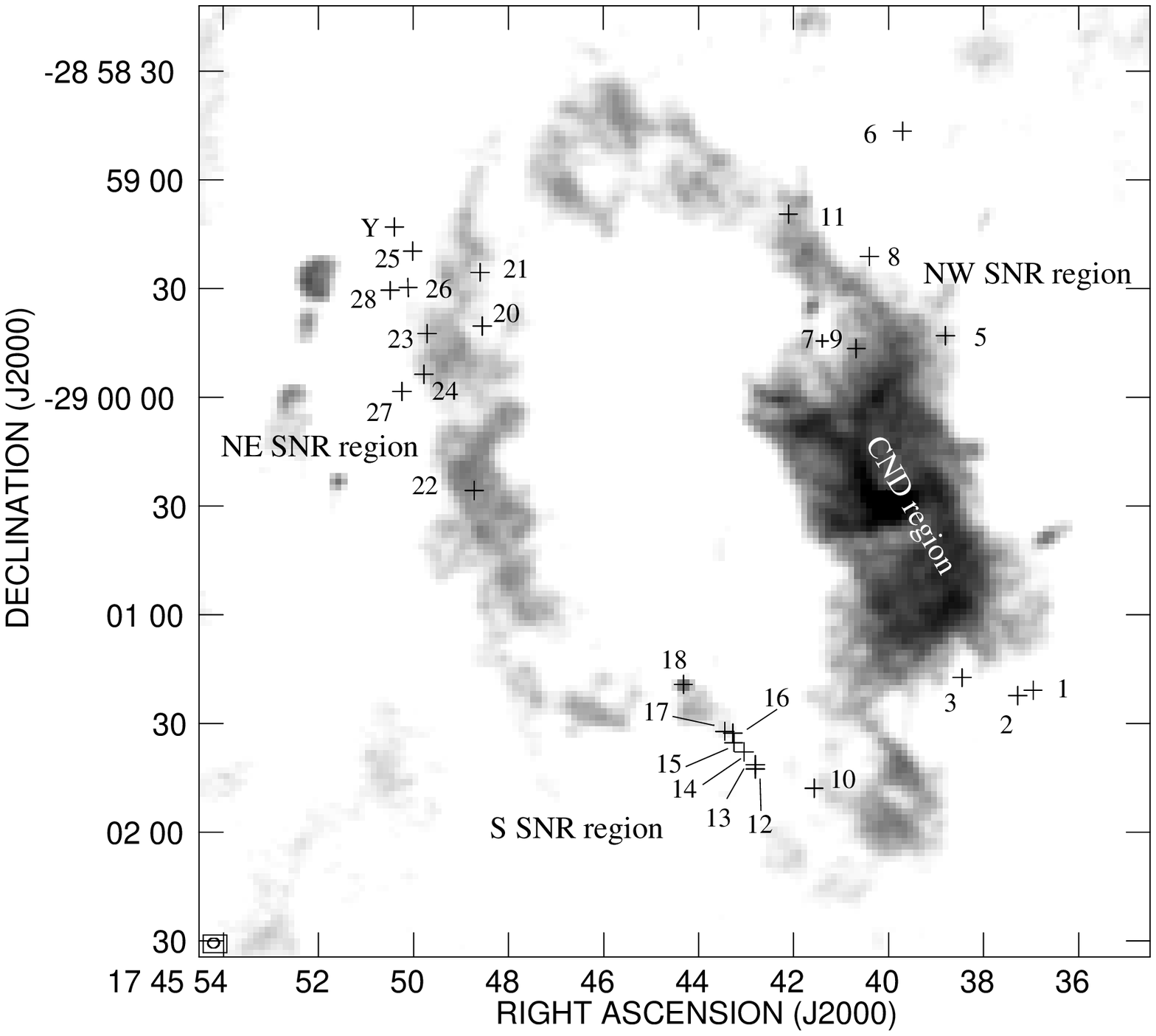}}}
\caption{Position of each maser listed in Table \ref{fluxes}, in
  relation to the 1.7\,GHz continuum.}
\label{maserpos}
\end{figure}

The most significant trend is that the region of SNR/ISM masers to the
northeast has a much larger fraction of variable masers than in other
regions. In the northeast, two out of 10 masers (80\%) have $VI>1.0$
with a median of $VI=1.77$. In contrast the southern SNR/ISM
interaction region, the three scattered northwestern masers and the
CND masers have two out of 15 masers (13\%) that are variable with a
median $VI=1.94$, while the majority of 13 masers (87\%) are
non-variable with a median $VI$ of 0.40. Figure \ref{maserpos}
displays the position of each maser feature, overlaid on the 1.7\,GHz
continuum flux density image. With this spatial distribution in mind,
the origin of the maser variability is discussed in Sect.\
\ref{discussion}.

\section{Discussion}
\label{discussion}
During the 195 days over which our study took place the most
significant result is that the northeastern masers associated with the
SNR/ISM interaction show a higher degree of variability than the rest
of the masers. Over the observed timescale no masers disappeared, and
no new masers appeared. Compared to our observations in 2005
(Paper\,I) one new maser appeared (or more precisely, this time
fulfilled our detection criteria). The two ``archival'' masers,
included for completeness in Paper\,I, remain undetected.
This indicates a quite stable environment for both SNR/ISM and CND
masers during our monitoring campaign. In this section we first discuss how
variability can be introduced by both external and
intrinsic (Sect.\ \ref{intr}) sources. Thereafter, we explore the
nature of the variability of the SNR (Sect.\ \ref{snrvar}) and CND
(Sect.\ \ref{cndvar}) masers respectively.

\subsection{Maser variability}
\label{intr}

Observational errors such as a systematic incorrect value in the
derived value of our flux calibrator in one of our epochs, pointing
errors, etc., will have a predictable signature in the individual
measurements. We have carefully looked for these in both our target
and calibrator data and conclude that none of them contribute to the
overall results. Flux variations can also be produced by interstellar
scintillation \citep{narayan92,cordes98}. This causes variations on
timescales of a few minutes. Such short timescales are not considered
in our study since in data averaged over a minute only a couple of
masers would be bright enough to be reliably detected.

Assuming that the variability of the maser intensity is not due to
observational effects, variability studies can be used to probe the
maser environment. Changes in maser peak velocity could indicate
acceleration, such as observed in outflows \citep{liljestrom89,
  brand03}. In stellar masers, periodic maser variations may be
related to periodicity found in the central object, in turn affecting
the pumping conditions and/or the maser path length. SiO masers in the
evolved Mira variables are examples of how the maser luminosity is a
result of the phase of the central star
\citep[e.g.][]{wittkowski07}. Another example can be found in star
forming regions (SFRs), where 22~GHz water masers are well-known to
exhibit maser variability. The variability of these masers are
characterized by large (sometimes several orders of magnitude)
amplitude variations, and often with velocity drifts of a few \kms\
per year \citep{comoretto90, wouterloot95, claussen96, brand03,
  furuya01, furuya03}. This occurs on timescales from a few hours to
years. Water masers associated with SFRs are excited behind shocks,
presumably caused by outflows or jets driven by the young stellar
object (YSO), and much of the variation can be attributed to changes
in the luminosity of the YSO, in turn affecting the pumping rate via
changes in the outflows and jets \citep{elitzur89, felli92}.

OH masers associated with SFRs are predominantly observed in the
radiatively pumped 1665 and 1667 MHz lines (sometimes accompanied by
masers in the 1612 and 1720\,MHz transitions), and variability of
those masers have been investigated by multiple groups
\citep{schwartz74, zuckerman72, rickard75, gruber76, clegg91}. The
long term variability (weeks to months) can be attributed to changes
in the number density of inverted OH molecules, or path length
changes. In contrast, the short term variability is assumed to be
related to sudden changes in the pumping mechanism, reflecting
fluctuations in the host star luminosity. For OH masers in SNRs, the
situation has been unknown. The 1720~MHz OH transition was for a long
time the only transition observed near SNRs \citep{lockett99,
  wardle99, pihlstrom08}, although recent observations have shown
Sgr\,A to harbor 36.2 and 44.1~GHz methanol masers
\citep{sjouwerman10,pihlstrom11}. The 1720~MHz OH masers are usually
assumed to originate in the post-shock region where the expanding SNR
collides with the surrounding medium. No dedicated variability studies
of 1720~MHz OH masers have been published to date, and we therefore
know little about the stability of the maser environment in these
sources. Here we have a shocked environment and since this is a large
scale phenomena it must arise in the maser generating column density,
or from the subsequent propagation of emission through the
interstellar medium.

\subsection{The SNR masers}
\label{snrvar}

The northeastern SNR masers clearly are variable on the timescales of
weeks to months. These masers are the result of the SNR Sgr\,A\,East
ramming into the $+50$~\kms\ cloud and we expect the post-shock medium
to display density variations due to turbulence in those regions,
given that a shock has recently passed. Assuming that the masers can
be considered as a series of velocity coherent regions of gas, bulk
gas flow and turbulent motion in the source will move parts of the
maser column in and out of velocity coherence along the line of
sight. As a result, the maser will exhibit intensity variations. For
the northeastern masers this cannot be the only source of the
variability though, since much higher levels of variability are
observed than toward other regions of Sgr\,A\,East. It is not clear
whether it is the maser generating column density itself or the medium
in front of the masers that is causing the variability. However, the
$+50$~\kms\ cloud is thought to be located behind Sgr\,A\,East
\citep[e.g.,][Paper\,I]{coil00} and thus between the observer and the
masers. When observing this group of masers we therefore probably look
through a longer path length of turbulent gas than toward other
regions of Sgr\,A\,East. This is supported by OH absorption
observations showing this region harbors a much higher absorbing OH
column density than the southern and northwestern interaction regions
(Paper\,I, Karlsson et al.\ 2003). It is possible that the supernova
covering the northeastern region contributes to the turbulence of this
column density.

In contrast, the southern masers are occurring in a region where the
two SNR Sgr\,A\,East and G359.02$-$0.09 are colliding. Here there is
no clear evidence of large columns of gas in front of the masers, and
no supernova crossing the line of sight.

\subsection{The CND masers}
\label{cndvar}

The CND masers exhibit much less variability than the northeastern SNR
masers, with no maser displaying variability with a
$VI>1$. If the northeastern SNR/ISM maser variability is due to
passage of the maser through a screen of hot turbulent gas, the 
(non-)variability of the CND masers is consistent with CND located
{\em in front of} Sgr\,A\,East (Paper\,I).

Since variability could give insight into the maser environment, we
consider whether the low variability is consistent with the presumed
pumping and amplification conditions in the CND. Here the pumping
cannot be due to a SNR/ISM shock, but the CND masers are thought to be
pumped by collisions between clumps in the CND (Paper\,I). The number
of clumps within the line of sight are probably only a few since once
a clump has moved out of the line of sight, a maser is likely to
disappear completely. Assuming an inclination of 65$^\circ$
\citep{latvakoski99,marr93,jackson93,christopher05} and thus a
rotational velocity of approximately 160~\kms\ of the maser emitting
clouds in the CND, the motion in the plane of the sky over the 195
days is minimal; only $3.7\times 10^{-5}$ pc (approximately 8
AU). This is much smaller than the high density cloud core sizes of
$\sim 0.25$ pc that are estimated by molecular line observations
\citep{christopher05}.  Similarly, assuming typical relative clump
velocities of 20-30~\kms\ in the sky-plane and 0.1~pc in size, a clump
will pass by a compact maser in $\sim$3900 years. For a clump size of
1~AU the corresponding time is 2 months. If clump collisions and
motions are involved in determining the path lengths for the CND
masers, given the low variability levels observed the clumps must be
much larger than 1~AU in size. More likely they are only a fraction of
the size of the dense molecular clumps observed in HCN, perhaps
corresponding to a high density region of the clump. For such clump
sizes little variability can be expected due to path length changes.

A factor that could contribute to a more stable maser gain in the CND
region is the pumping conditions. As measured in molecular line
emission, HCN and HCO$^+$ line intensity is stronger in the CND than
in regions associated with, for example, the $+50$~\kms\ cloud
\citep{wright01}, implying higher densities. A higher density would
result in a higher collision rate, thus keeping the pumping at a
higher rate eventually resulting in more stable maser intensities.
Note that \cite{yusefzadeh01} show that densities $n>10^6$ cm$^{-3}$
are needed for clump-clump collisions to produce 1720\,MHz pumping,
which is slightly higher than the C-shock post-region density of
$n\simeq10^5$ cm$^{-3}$ for the SNR masers \citep{lockett99,wardle99}.

What the exact pumping conditions and the pumping rate really are is
difficult to calculate, and depends on the ortho-para $H_2$ ratio in
the region. Both \citet{lockett99} and \citet{pavlakis96} have found
that collisions with para-H$_2$ can strongly suppress the 1720 MHz
inversion. Ortho- and para-H$_2$ are thought to be formed on grains
with a ratio of 3:1.  At low temperatures ($T=10$ K) proton exchange
reaction convert ortho-H$_2$ into para-H$_2$, making the para-H$_2$
being the dominant species. As the temperature increases, there is
less of a difference, but still with para-H$_2$ dominating
\citep{offer92}. However, to create the 1720~MHz maser only a small
amount of ortho-H$_2$ is required, since the collision rate for
ortho-H$_2$ usually is larger by a factor of 2-3 than the para rates
\citep{offer92}. The equilibrium ratio for ortho-para H$_2$ is
$9\times e^{-170/T}$, which is equal to 1 for a temperature $T=77$\,K,
implying the temperature should be cooler than this value. This is
consistent with a high density environment in the CND where the gas
temperature is thermalized with the dust temperatures of 20-80\,K
\citep{becklin82,mezger89}, although molecular studies of the CND
imply excitation temperatures anywhere between 50-200\,K
\citep{christopher05,jackson93,wright01,coil99,coil00}. \citet{coil00}
use NH$_3$(2,2)/NH$_3$)(1,1) ratios to derive rotational gas
temperatures between 20-70~K, mostly in the outer regions of the
CND. Perhaps, next to path-length geometry (Paper\,I), this is a
reason that 1720 MHz masers are not seen all across the CND; if
temperatures are too high the 1720 MHz emission will be suppressed.

\section{Summary}
We have presented the first flux monitoring study of 1720~MHz OH
masers associated with an SNR. In total five epochs of the 1720~MHz OH
masers in the GC were obtained, and used to estimate the maser flux in
different regions of the GC. We find that the typical variability is
very low, but that the northeastern SNR/ISM masers have a higher
variability than the other masers. We speculate this is because the
$+50$~\kms\, cloud is located behind the SNR, and the maser emission
travels through a longer path length of turbulent gas before reaching
the observer. At other locations across the SNR the masers probably
arise at the near side of the SNR with less material between the maser
and the observer, and therefore show a much lower level of
variability. Similarly low variability levels are found for the CND
masers consistent with the CND being located in front of
Sgr\,A\,East. The low variability of the CND maser agrees with a
scenario where the pumping is provided by relatively large clumps
(sizes $>1$ AU).

\acknowledgments The National Radio Astronomy Observatory is a
facility of the National Science Foundation operated under cooperative
agreement by Associated Universities, Inc.

{\it Facilities:} \facility{VLA}

\end{document}